**Opposite current-induced spin polarizations in bulk-metallic Bi$_2$Se$_3$ and bulk-insulating Bi$_2$Te$_2$Se topological insulator thin flakes**

Jifa Tian[1,2,3,*], Cüneyt Şahin[4], Ireneusz Miotkowski[1], Michael E. Flatté[4], and Yong P. Chen[1,2,5,6,7,*]

1. Department of Physics and Astronomy, Purdue University, West Lafayette, IN 47907, USA

2. Birck Nanotechnology Center, Purdue University, West Lafayette, IN 47907, USA

3. Department of Physics and Astronomy, University of Wyoming, Laramie, WY, 82071, USA

4. Optical Science and Technology Center and Department of Physics and Astronomy, University of Iowa, Iowa City, Iowa 52242, USA

5. Purdue Quantum Science and Engineering Institute, Purdue University, West Lafayette, IN 47907, USA

6. School of Electrical and Computer Engineering, Purdue University, West Lafayette, IN 47907, USA

7. Institute of Physics and Astronomy and Villum Centers for Dirac Materials and for Hybrid Quantum Materials, Aarhus University, 8000 Aarhus-C, Denmark

*Corresponding authors: jtian@uwyo.edu; yongchen@purdue.edu

**Abstract**

One of the most fundamental and exotic properties of 3D topological insulators (TIs) is spin-momentum-locking (SML) of their topological surface states (TSSs), promising for potential applications in future spintronics. However, other possible conduction channels, such as a trivial two-dimensional electron gas (2DEG) with strong Rashba type spin-orbit interaction (SOI) and bulk conducting states that may possess a spin Hall effect (SHE), can coexist in 3D TIs, making determining the origin of the current induced spin polarization (CISP) difficult. In this work, we directly compared the CISP between bulk-insulating Bi$_2$Te$_2$Se (BTS221) and bulk-metallic Bi$_2$Se$_3$ thin flakes using spin potentiometry. In the bulk insulating BTS221, the observed CISP has a sign consistent with the expected helicity of the SML of the TSS, but an opposite sign to its calculated bulk spin Hall conductivity (SHC). However, compared to BTS221, an opposite CISP is observed in the bulk metallic Bi$_2$Se$_3$, consistent instead with both the expectations of its Rashba-Edelstein effect of the band-bending induced 2DEG and bulk spin Hall Effect (SHE). Our results provide an electrical way to distinguish the TSS from other possible conducting channels in spin transport measurements on 3D TIs, and open ways for the potential applications in charge-spin conversion devices.

## I. INTRODUCTION

Charge carriers in materials with strong spin-orbit interaction (SOI) commonly experience a momentum-dependent effective magnetic field and a geometric phase.[1-3] These features are particularly attractive for the realization of device concepts in which spin polarization is generated from a charge current, manipulated by electric fields and detected electrically or optically.[3] Three-dimensional topological insulators (3D TIs) possess nontrivial spin-momentum-locked topological surface states (TSSs) representing a new type of the SOI. The charge carriers of TSS are massless Dirac fermions with their spins locked in-plane and perpendicular to their momentum (Fig. 1a),[4-10] making the TI system a promising source for generating spin polarization and spin current without ferromagnetic (FM) materials. However, real 3D TI materials can also host trivial 2D electron gas (2DEG) with strong Rashba type SOI near the surfaces derived from their bulk states (Fig. 1a).[11,12] These Rashba 2DEGs typically have two Fermi surfaces (Fig. 1a) with opposite spin helicities, where the outer one (which contributes more to the transport than the inner one[13]) has the opposite spin helicity compared to that of TSS[11,12]. Furthermore, the bulk of many 3D TIs may be doped conducting, and can possess a spin Hall effect (SHE, Fig. 1a) resulting from the heavy elements present. In recent years, spin potentiometry has been applied to study current induced spin polarization (CISP) in 3D TIs, where the direction of spin polarization can be determined from a voltage difference (thus a spin chemical potential difference) measured between opposite magnetizations of an FM tunneling probe, controlled by an in-plane magnetic field applied along its easy axis.[14-21] However, so far, most of the reported CISPs measured on different 3D TIs have been solely ascribed to TSS, even though some of the spin signals (voltage differences) show opposite signs[14-22]. Especially, the contributions to CISP from Rashba 2DEG and bulk intrinsic SHE in 3D TIs have been largely ignored. Thus, to clarify the origins of the CISP in 3D TIs, a careful study considering all the possible contributions is desired.

In this work, we systematically studied and compared the CISPs in bulk insulating $Bi_2Te_2Se$ (BTS221) and bulk metallic $Bi_2Se_3$ TI thin flakes by spin potentiometry. The spin devices (Fig. 1b,c) of the two materials were made using a similar nanofabrication process and measured in the same system. The FM tunneling probe detects the CISP on the *top* surface. In both materials, we observed a similar step-like spin signal when an in-plane magnetic field is applied to switch the magnetization of the FM voltage probe (the spin detector), and reversing the current direction reverses the sign of the spin signal. The CISP in bulk insulating BTS221 is found to be consistent with the theoretical expectations of TSS, whereas the CISP measured in bulk metallic $Bi_2Se_3$ is opposite, and is consistent with the Rashba-Edelstein effect from the band bending induced 2DEG near the top surface. We further studied the back gating, current, and temperature dependences of the spin signals measured in these two materials. Furthermore, we calculated the bulk intrinsic SHC of $Bi_2Se_3$ and BTS221, and discussed its possible contribution to the observed CISP. Our work clarifies and highlights the differences in spin transport from TSS, Rashba 2DEG, and bulk SHE, and enables the potential applications in spintronics by utilizing these states in 3D TIs.

## II. EXPERIMENTAL METHODS

High-quality BTS221 and $Bi_2Se_3$ single crystals were grown by the Bridgman method.[18,23,24] The 3D TI thin flakes (with a typical thickness of ~10-40 nm, and lateral size up to ~5-10 μm) were exfoliated from these bulk crystals and placed on top of $SiO_2$ (thickness 285 nm)/Si (heavily doped) substrates serving as the back gate. The BTS221 crystal is bulk insulating [18], while the $Bi_2Se_3$ crystal is heavily n-type doped (presumably due to the Se vacancies [23]). The spin devices of the two materials were fabricated using a similar process as reported in our previous work [18]. For fabricating the spin devices, we used both the permalloy (Py, $Ni_{0.80}Fe_{0.20}$) and cobalt (Co) as the FM probes (the spin potentiometer), separated from the TI surface by one nm-thick $Al_2O_3$. The non-magnetic contacts are Au (Figs. 1b,c and 2a,d).

Two circuit layouts (Hall bar-like and four-terminal configurations, schematically shown in Fig. 1b,c, respectively) were used in the spin potentiometric measurements on both the BTS221 and $Bi_2Se_3$ devices. A dc current was applied between two "outmost" Au electrodes and a voltage ($V$) was measured between an FM (Co or Py) and the adjacent Au contacts using a high-impedance voltmeter. We define a positive current ($+I$) as flowing from right to left along $-x$ direction (thus the electrons flow from left to right along $+x$ direction), and a positive in-plane magnetic field ($+B$) as applied in the $-y$ direction along the easy axis of the FM electrode (so its majority spin points to $+y$ direction)[18]. A voltage difference (spin signal) $\delta V$ can be extracted when the magnetization of the FM tunneling probe is reversed. We note that the sign of the spin signal and the CISP for a given current are independent on the choices of the circuit layout and the FM contacts. All measurements were performed in a variable temperature cryostat with a base temperature ($T$) of 1.5 K.

## III. RESULTS AND DISCUSSIONS
### A. CISP in BTS221 and $Bi_2Se_3$ 3D TI samples

We directly compared the spin transport results measured on a 40-nm-thick BTS221 (device A, Fig. 2a) and a 35-nm-thick $Bi_2Se_3$ (device B, Fig. 2d). For the bulk insulating BTS221 sample, the voltages ($\Delta V$) measured between a Py spin detector (contact "2" in device A) and an adjacent Au contact ("3") as a function of the in-plane magnetic field ($B$) are detailed in Figs. 2b,c, where a linear background has been subtracted (Fig. S1 in the Supplemental Material [25]). We see a high (low) voltage state when the channel spin polarization $s$ is parallel (antiparallel) to the direction of the magnetization $M$ of the FM contact (equivalently, when the FM majority spins, which are oriented opposite to $M$, are antiparallel (parallel) to the channel spin polarization *s*, the FM contact measures the lower (higher) electrical chemical potential μ of the less (more) occupied channel spin states). We note that the voltage V has a direct relationship with the electrical chemical potential μ of V = μ/*q*, where charge *q* = -*e* if the charge carriers are electrons. The sign of the measured spin signal δV (=ΔV$_{+M}$ - ΔV$_{-M}$) can be extracted and used to determine the direction of CISP.[18,22] As shown in Fig. 2b, when a positive bias current *I* is applied along the -x direction and the magnetization *M* of the Py detector is in -y (+y) direction, we observed a high (low) voltage state with a spin signal δV ~20 μV. Thus, we can determine that the direction of the spin polarization *s* in the bulk insulating BTS221 flake is along the -y direction (red arrow) for a positive *I* along the -x direction (electron momentum along +x direction), such that a +y-magnetized detector (majority spins along -y) would detect a high occupation or chemical potential (thus low voltage). As expected, reversing the bias current (*I* = -75 μA) reverses the sign of the spin signal (now δV ~ -20 μV in Fig. 2c) and the corresponding channel spin polarization *s*. From the above analyses, we conclude that the CISP in the bulk-insulating BTS221 is consistent with the expectation of TSS due to SML [18,19,21,26]. In stark contrast, for the bulk metallic Bi$_2$Se$_3$ (Fig. 2d, device B), a low (high) voltage state is observed at the positive (negative) magnetization *M* long -y (+y) and a positive *I* = 10 μA along -x (Fig. 2e,), showing an opposite trend of the hysteretic step-like spin signal (δV < 0) comparing to that of BTS221. Reversing current *I* to -10 μA reverses the trend of the spin signal (δV >0) (Fig. 2f). Given the same sign of the bias current *I* (Fig. 2), the opposite spin signal observed in Bi$_2$Se$_3$ compared to that of BTS221 indicates an opposite CISP (the red arrows in Fig. 2) and a non-TSS origin. We note that the measured spin signal and the direction of the CISP in bulk-metallic Bi$_2$Se$_3$ samples are consistent with that of the MBE-grown Bi$_2$Se$_3$ previously reported [14]. As demonstrated by the spin-resolved angle-resolved photoemission spectroscopy (ARPES) [11,12], the spin helicity of the outer Fermi surface of the Rashba 2DEG in Bi$_2$Se$_3$ is opposite to that of TSS, suggesting that the Rashba 2DEG [13] can be one of the causes for the observed opposite CISP in Bi$_2$Se$_3$. In addition, we will also discuss possible contribution of bulk spin Hall effect as another possible cause in Section D.

### B. Current, temperature and gate dependences of the spin signals for BTS221

We studied the current and temperature dependences of the spin signals for BTS221 (device A). Figure 3a shows the voltage measured by the Py spin detector as a function of the in-plane *B* field at different bias currents ranging from ±0.5 μA to ± 100 μA at *T* = 1.6 K. To highlight the current dependence of the spin signal, all of the curves have been vertically offset to have their central values aligned. The corresponding spin signal δV as a function of the bias current *I* is summarized in Fig. 3b. We see that δV linearly increases with the applied current *I*, consistent with the expectations of CISP [13]. The temperature dependent spin signals were measured at bias currents *I* = ±100 μA. The amplitude of the spin signal |δV| shows a weak dependence on temperature when *T* is below 60 K. However, when *T* is above 60 K, |δV| decreases steadily

and vanishes at ~120 K. After comparing the temperature dependence of the spin signal with that of the sample resistance $R$ (Fig. S2 in the Supplemental Material [25]), we find that, similar to the temperature-dependent spin signal, $R$ changes its trend at ~ $T$ = 60 K, suggesting that the bulk conduction becomes notable when $T$ is > 60 K. We note that the reduction and disappearance of the spin signal at higher temperatures may be attributed to various factors, such as the increased contributions from the bulk states, increased scattering (due to phonons) and thermal fluctuations of the polarized spins [16], and degradation of the quality of $Al_2O_3$ tunneling barriers [21], etc.

We further studied the gate dependence of the spin signal measured on the bulk-insulating BTS221 sample (Fig. 4). The representative spin signals measured on device A at different gate voltages ($V_g$) are shown in Fig. 4a, where the spin results measured at a bias current of +20 μA (-20 μA) are shown in the top (bottom) panel. We note that all the voltage traces have been vertically offset to have their central values aligned. The gate dependence of the spin signal δV is summarized in Fig. 4b, where the spin signal |δV| is enhanced by ~ 3 times as $V_g$ is tuned from 80 V to -80 V. Such an enhancement is consistent with the gate-dependence of the voltage ($V_0$) measured at $B$ = 0 T and a bias current of $I$ = 20 μA as shown in Fig. 4c, where $V_0$ can be tuned with an on-off-ratio of ~ 4, confirming a strong back gate effect of our BTS221 device. We further note that the qualitiative trends of the gate-dependent spin signal (δV increases as the Fermi level is tuned to approach the Dirac point) as well as the CISP are consistent with the experimental results measured on bulk insulating $(Bi,Sb)_2Te_3$ thin films grown on the $SrTiO_3$ substrates where an ambipolar field effect has been achieved [21].

### C. Current and temperature dependences of the spin signals for $Bi_2Se_3$

In order to have a better understanding of the opposite spin helicity observed in the bulk metallic $Bi_2Se_3$, we systematically studied the current and temperature dependences of the spin signal. Figure 5a shows the spin signal (measured on device B) as a function of the in-plane magnetic field at different bias currents ranging from ± 0.05 μA to ± 100 μA at $T$ = 1.6 K. The extracted spin signal δV (negative for +$I$ and positive for -$I$) as a function of $I$ is summarized in Fig. 5b, where |δV| linearly increases with the increasing bias current at |$I$| < 20 μA but saturates (possibly due to Joule heating) at high currents (|$I$| > 20 μA). The temperature dependence of the spin signals measured at $I$ = ±15 μA is summarized in Fig. 5c. We see that, different from BTS221, as the temperature increases, the spin signal δV of $Bi_2Se_3$ quickly decreases and disappears at ~80 K. Other than the above listed possible factors for BTS221, the quick reduction of the spin signal in $Bi_2Se_3$ can be consistent with the expectations of Rashba-Edelstein effect from the 2DEG [13] since its CISP reduces much faster than that of TSS as temperature increases (possibly reflecting the relative proximity to other states that can be thermally excited).

### D. Origin of the CISP in BTS221 and Bi$_2$Se$_3$

Depending on the location of the Fermi level, different electronic states and mechanisms, such as TSS, Rashba 2DEG, and the bulk SHE, can contribute to the measured CISP in 3D TIs. We can infer that the Fermi level in BTS221 is located in the TSS based on its temperature dependence of the resistance R (Fig. S2a in the Supplemental Material [25], exhibiting a weakly-insulating behavior for T > 60 K) and the extracted (including carriers from both the top and bottom surfaces) 2D carrier density of ~ 1×10$^{13}$ cm$^{-2}$ (~5×10$^{12}$ cm$^{-2}$ for one surface, Fig. S2 in the Supplemental Material [25], because the maximum carrier density of TSS at the bottom of the bulk conduction band for *each* surface in BTS221 is ~ 1×10$^{13}$ cm$^{-2}$ calculated based on the band structure measured by ARPES[24]). Together with their current and temperature dependences of the spin signals (Fig. 3) and its calculated bulk intrinsic SHC (Fig. 6b, discussed below), we can conclude that the observed CISP in our bulk insulating BTS221 is mainly due to SML of TSS [18,21]. As for Bi$_2$Se$_3$, its Fermi level is located in the bulk conduction band [27] because of its clearly metallic behavior (Fig. S2d in the Supplemental Material [25]) and the high 2D carrier density (~7×10$^{13}$ cm$^{-2}$, ~ 7 times higher than that of BTS221, Fig. S2f in the Supplemental Material [25], since the maximum carrier density of TSS for each surface in Bi$_2$Se$_3$ is ~ 8.9×10$^{12}$ cm$^{-2}$ calculated based on the ARPES-measured band structure[28]). Thus, the measured CISP (with opposite sign compared to that of BTS221) has little contribution from the SML of TSS, but comes from other possible states with opposite spin helicity, such as the Rashba-Edelstein effect from the band-bending induced 2DEG near the top surface, where its outer Fermi circle features an opposite spin helicity (demonstrated by the spin-resolved ARPES[11,12]) compared to that of TSS.

In addition to SML of TSS and Rashba-Edelstein effect of the Rashba-type 2DEG as the possible causes for the CISP in 3D TIs, here we discuss another possible source arising from the bulk intrinsic SHE [29] of Bi$_2$Se$_3$ and BTS221. A tight-binding Hamiltonian consisting of *s*- and *p*- orbitals of the Bi, Te and Se atoms has been constructed, including SOI with parameters taken from an earlier study [30]. We calculated the intrinsic SHC of the Bi$_2$Se$_3$ and BTS221 crystals in terms of a spin current-charge current response function within the linear response theory using the Kubo formula.[31] The SHC is the sum of the Berry curvatures of the filled bands up to the Fermi level. We computed the energy-dependent density of Berry curvatures which allows us to determine the SHC, $\sigma_{ij}^k$, as a function of the chemical potential, where i, j, and k stand for the direction of the spin current, the direction of the charge current, and polarization of the spin, respectively. The coordinates used in the calculation are shown in Fig. 6a. To directly compare the theoretical and experimental results, we chose the z, x, and y axes as the directions of spin current, charge current, and spin polarization. We summarized the calculated SHC, $\sigma_{zx}^y$, as a function of Fermi energy $E_F$ in Fig. 6b. We do not observe a significant anisotropy in the SHC results for different directions, especially for Fermi levels around the band gap. This can be attributed to the rhombohedral crystal structure of these materials, which is closely related to a simple cubic structure by a slight distortion along the body diagonal. We can further determine the direction of CISP from the sign of the calculated SHC due to the intrinsic bulk SHE. As shown in Fig. 6b, the negative SHC $\sigma_{ij}^k$ indicates that if a charge current (electron current) flows along the +x (-x) axis, the spin current will flow to the +z axis and the induced spin polarization will point to -y axis. Based on this analysis, we can conclude that the CISP due to bulk intrinsic SHE for both $Bi_2Se_3$ and BTS221 would have the same sign, which would be the same as that of Rashba-Edelstein effect from the 2DEG (possibly observed in $Bi_2Se_3$), but opposite to that of TSS (as observed in BTS221). Thus, in $Bi_2Se_3$ both the intrinsic SHE and band bending induced Rashba 2DEG can be the causes of the observed opposite CISP compared to that of BTS221. In addition to the bulk intrinsic SHE, in $Bi_2Se_3$, the extrinsic SHE due to the spin-dependent scattering (known as skew-scattering) around defects (e.g., Se vacancies) may also exist. Since these Se vacancies normally create positively charged defects, electrons scattered to the top surface would have the same spin-polarization direction as that of TSS,[32] resulting in the opposite sign of the spin signals measured by spin potentiometry. Thus, the skew-scattering induced SHE appears unlikely to be the cause of our observed spin signal in $Bi_2Se_3$. We note a recent work on measurements of the SHC in Cr-doped $(Bi,Sb)_2Te_3$ thin films using spin pumping [33] on a series of samples whose Fermi energy can be largely tuned to the valence band, the bulk energy gap, and the conduction band. We note prior experiments show that the CISP in $(Bi,Sb)_2Te_3$ thin films have the same sign as what we observe in BTS221 [18, 21, 22]. The sign of the calculated bulk intrinsic SHC of $(Bi,Sb)_2Te_3$ was found to be consistent with the spin pumping signal measured in Cr-doped $(Bi,Sb)_2Te_3$ samples [33]. The sign of the calculated SHC of $(Bi,Sb)_2Te_3$ is consistent with our calculated SHC of $Bi_2Se_3$ and BTS221 if the same coordinate system is used. On the other hand, the facts that samples in Ref. 33 are doped with chromium and have a different composition than ours and the measurement technique is also different from ours make a proper comparison of the experiments challenging.

A previous report pointed out that a local Hall effect due to the fringe field of an FM detector could also induce a hysteretic step-like voltage change, which was observed in MBE-grown $Bi_2Se_3$ samples [27]. In our case, we have fabricated square-shaped FM detectors on both BTS221 and $Bi_2Se_3$ samples (Figs. S3,4). The representative CISP can only be detected when the applied bias current is perpendicular to the in-plane magnetic field, whereas no step-like spin signal (Figs. S3,4) is observed when the bias current is parallel to the magnetic field. Our results are fundamentally different from the reported fringe field-induced step-like voltage changes [27]. Furthermore, the opposite signs of the spin signals (Fig. 2) in BTS221 and $Bi_2Se_3$ cannot be explained by the fringe field-induced Hall effect because both samples show n-type (which should give the same sign of the fringe field-induced Hall signal) charge carriers.

## IV. CONCLUSIONS

In conclusion, we have directly compared the CISPs in bulk insulating BTS221 and bulk metallic $Bi_2Se_3$ thin flakes. We find that, for a given bias current and magnetization of the FM spin detector, the measured sign of the spin signal and the direction of the CISP in BTS221 are opposite to that of $Bi_2Se_3$. The CISP observed in our bulk insulating BTS221 can be unambiguously ascribed to TSS due to SML. The opposite spin polarization observed in our bulk metallic $Bi_2Se_3$ can be due to the band-bending induced Rashba 2DEG as well as the bulk intrinsic SHE. Our results systematically studied the roles of different states in the observed CISP in 3D TIs, and open a potential way to manipulate spins by utilizing SML of TSS, Rashba 2DEG, and intrinsic bulk SHE for novel 3D TI-based spin and nanoelectronic devices.

**ACKNOWLEDGMENTS**

This work was partially supported by DARPA MESO program under Award N66001-11-1-4107 and NSF under Award EFMA-1641101. J. T. also acknowledges DOE, Office of Basic Energy Sciences, Division of Materials Sciences and Engineering for financial support under Award DE10SC0020074.

[33]  H. Wang, J. Kally, C. Şahin, T. Liu, W. Yanez, E. J. Kamp, A. Richardella, M. Wu, M. E. Flatté, and N. Samarth, Fermi level dependent spin pumping from a magnetic insulator into a topological insulator, Phys. Rev. Research **1**, 012014 (2019).

**Figure Caption**

**Figure 1** Schematic illustrations of possible states that could give current induced spin polarization on surfaces of 3D topological insulator (TI) materials and the circuit layout used in the spin potentiometric measurements.

(a) Schematic illustration of topological surface state (TSS), band-bending-induced Rashba-type two-dimensional electron gas (2DEG) near the top surface, and bulk spin Hall effect (SHE) in 3D TIs. The red, black, and blue spheres with arrows representing the current-induced spin polarization due to TSS, 2DEG, and bulk SHE, respectively. The schematics of the Fermi surface of TSS (in red) and Rashba 2DEG (in gray) are shown in (a). Schematic device structures and circuit layouts used in the spin potentiometric measurements: (b) Hall-probe like and (c) four-terminal geometries. The TI surface defines the x-y plane and the surface normal z direction. The two outside nonmagnetic (e.g., Au) contacts are used to apply a dc bias current (*I*), and the middle ferromagnetic (e.g., Py or Co) contact is magnetized by an in-plane magnetic field *B* along its easy axis (*y*, with -*y* defining the positive *B* and positive *M* directions).

**Figure 2 Current-induced spin polarizations with opposite signs in bulk-insulating $Bi_2Te_2Se$ (BTS221) and bulk-metallic $Bi_2Se_3$ thin flakes.**

(a) Optical image of a typical BTS221 device (device A, a 40 nm-thick flake). The corresponding circuit layout is shown in Fig. 1b. (b,c) Voltage ($\Delta V$) measured by an FM (Py, electrode "2") spin detector as a function of the in-plane magnetic field (*B*) at bias currents *I*= 75 μA (b), and -75 μA (c), where a linear background has been subtracted from all the traces. The purple arrow represents the extracted spin signal $\delta V = \Delta V_{+M} - \Delta V_{-M}$. Here, $\Delta V_{+M}$ and $\Delta V_{-M}$ are the representative voltages measured before and after the magnetization switching of the FM spin detector near the coercive field. Insets of b,c are the schematic band structures of TSS under positive and negative bias currents, respectively. (d) Optical image of a representative $Bi_2Se_3$ device (device B, a 35 nm-thick flake). (e,f) Voltage ($\Delta V$) measured by a Co (electrode "2") spin detector as a function of the in-plane *B* field at bias currents *I*= 10 μA (e), and -10 μA (f). The raw data of the voltage traces are vertically shifted to be centered at zero. For (b,c, e,f), the directions of the bias current *I*, current-induced spin polarization (CISP) *s* at the top surface (inferred from the sign of the spin signal) and magnetization *M* of the FM are labeled by the corresponding arrows. The arrows on the data traces indicate magnet field sweep directions. All the measurements were performed at T = 1.6 K.

**Figure 3 Current and temperature dependences of the spin signal measured on bulk insulating BTS221.**

(a) The voltage measured by the Py spin detector on device *A* as a function of the in-plane magnetic field at different bias currents. The voltage traces have been vertically offset to have their central values aligned. (b) The spin signal δV as a function of the applied bias current. All the measurements are performed at T = 1.6 K. (c) The spin signal δV as a function of temperature ranging from 1.6 K to 125 K measured at a bias current of +100 µA or -100 µA.

**Figure 4 Gate dependence of the spin signal measured on the bulk insulating BTS221.** (a) The voltage measured by the Py spin detector on device *A* as a function of the in-plane magnetic field at different back gate voltages ($V_g$) and bias currents of 20 µA (top panels) and -20 µA (bottom panels), respectively. The voltage traces have been vertically offset to have their central values aligned. (b) Spin signal δV as a function of $V_g$ for both positive (red squares) and negative (black triangles) bias currents. (c) Back-gate dependence of the voltage ($V_0$) measured at zero *B* field at a bias current of *I* = 20 µA. Insets are the schematics of the band structure of TI with the corresponding Fermi levels at most positive and negative gate voltages. All the measurements were performed at T = 1.6 K.

**Figure 5 Current and temperature dependences of the spin signal measured on bulk metallic Bi$_2$Se$_3$.** (a) The voltage measured by a Co spin detector on device *B* as a function of the in-plane magnetic field at different bias currents, where the voltage traces have been vertically offset to have their central values aligned. (b) The spin signals δV as a function of the applied bias current *I*. All the measurements were performed at T = 1.6 K. (c) The spin signal δV as a function of the temperature ranging from 1.6 K to 80 K.

**Figure 6 The calculated intrinsic spin Hall conductivity (SHC) in the bulk of Bi$_2$Se$_3$ and BTS221.** (a) The coordinate system based on the crystal structure of Bi$_2$Se$_3$ and BTS221 used in the calculations. (b) The intrinsic SHC $\sigma_{ij}^k$ of Bi$_2$Se$_3$ and BTS221 crystals as a function of the Fermi energy ($E_F$). Here, we consider i, j and k along the z, x, and y axes as the directions of spin current, charge current, and spin polarization, respectively.

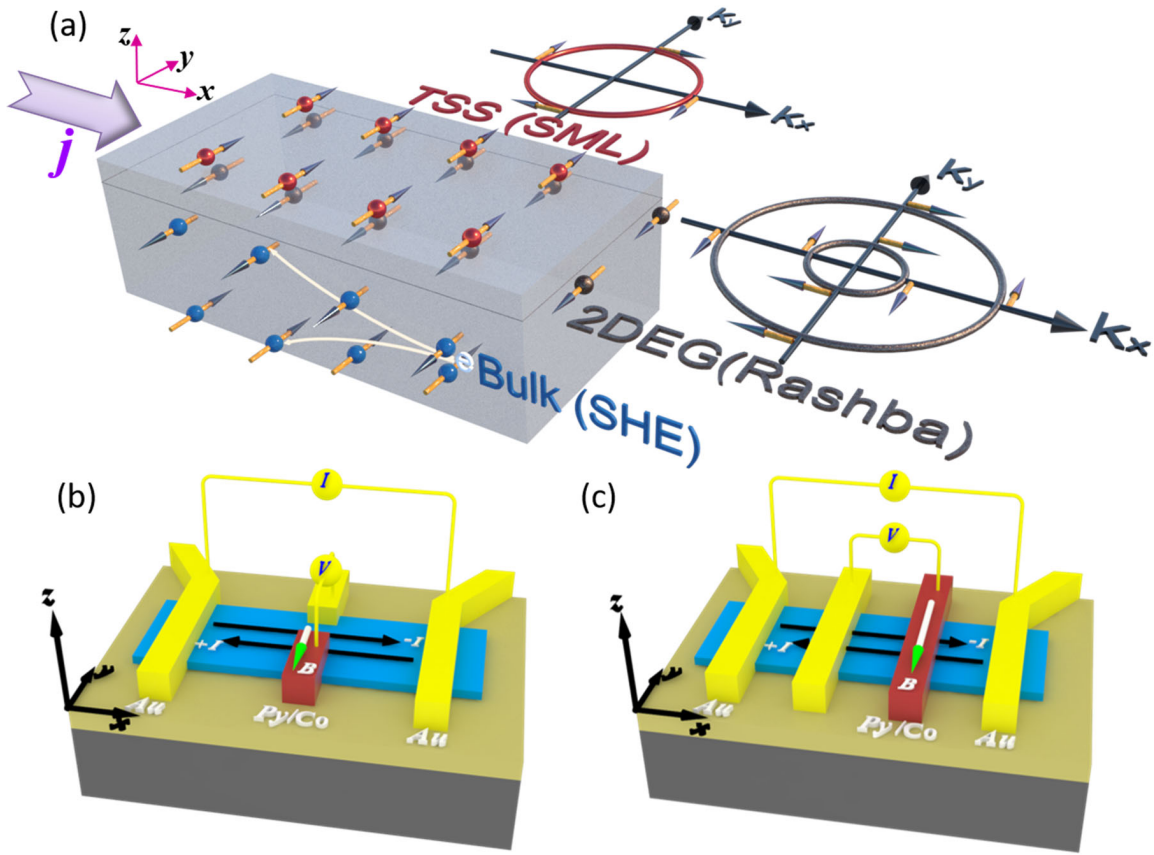

Figure 1 by Tian et al.

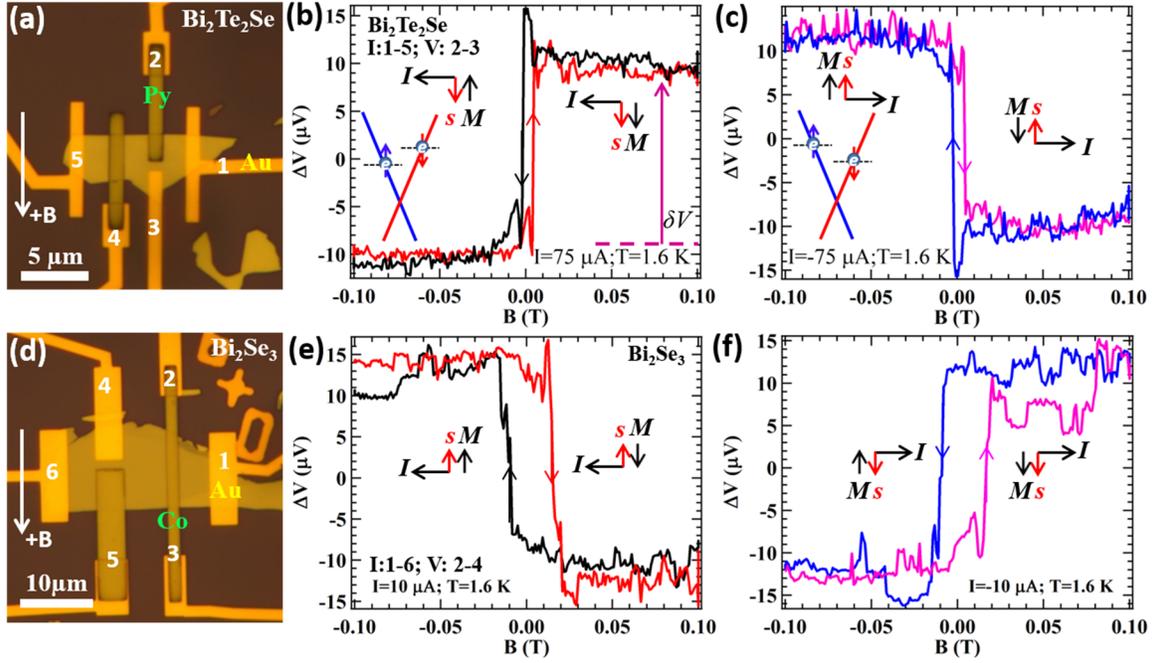

Figure 2 by Tian et al.

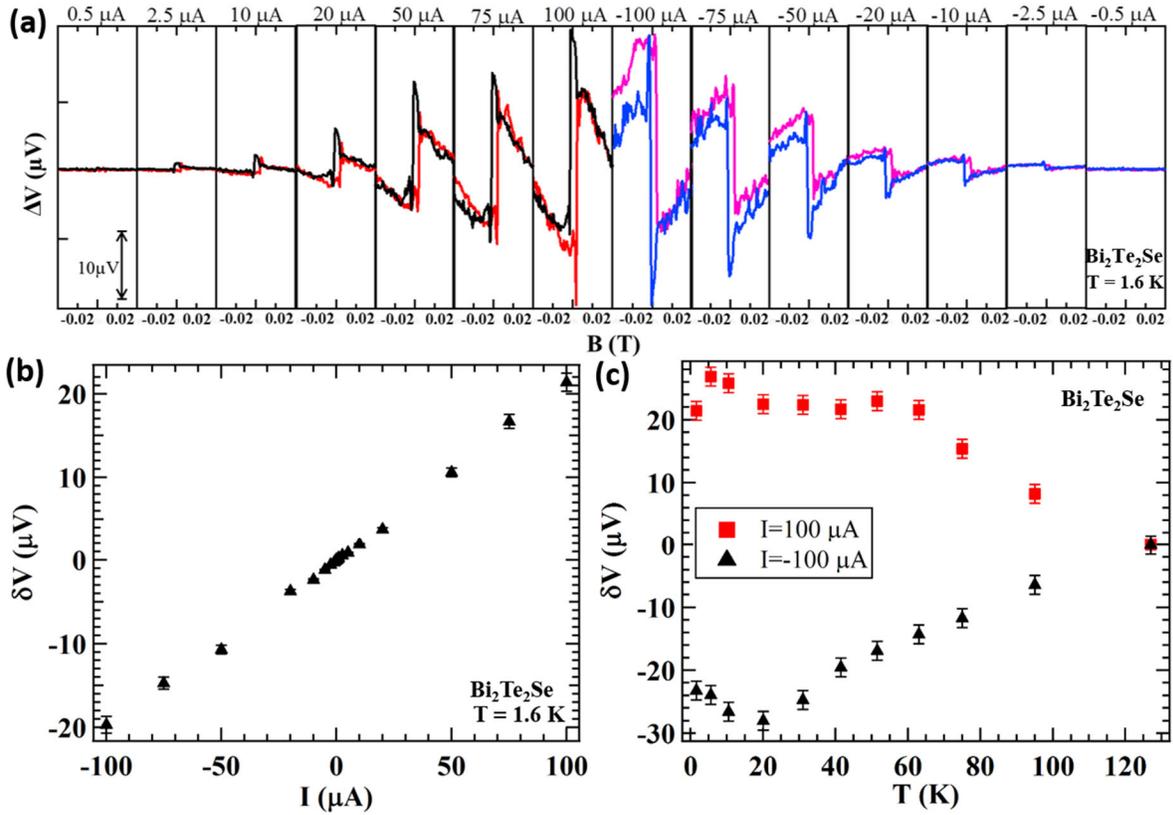

Figure 3 by Tian et al.

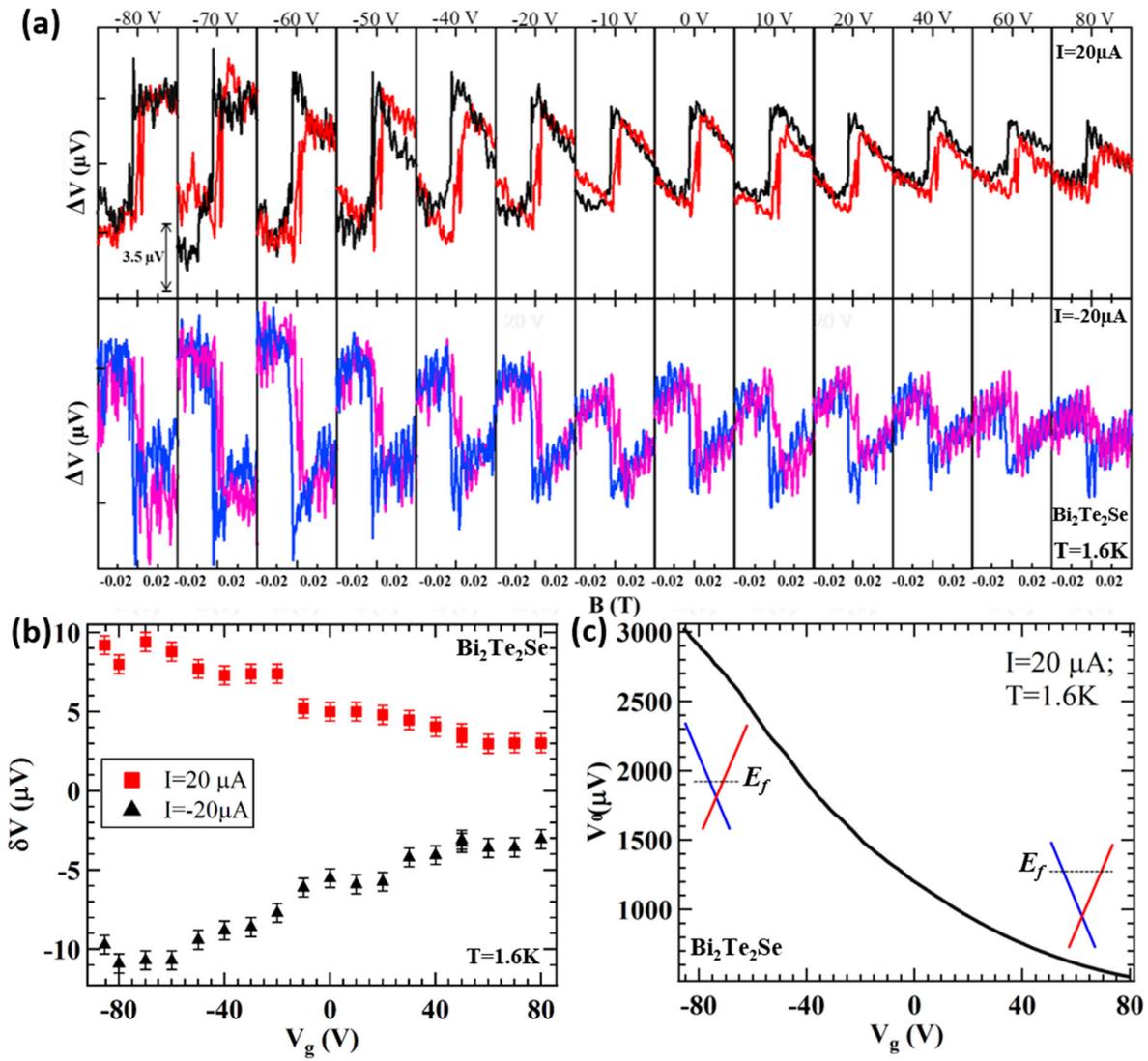

Figure 4 by Tian et al.

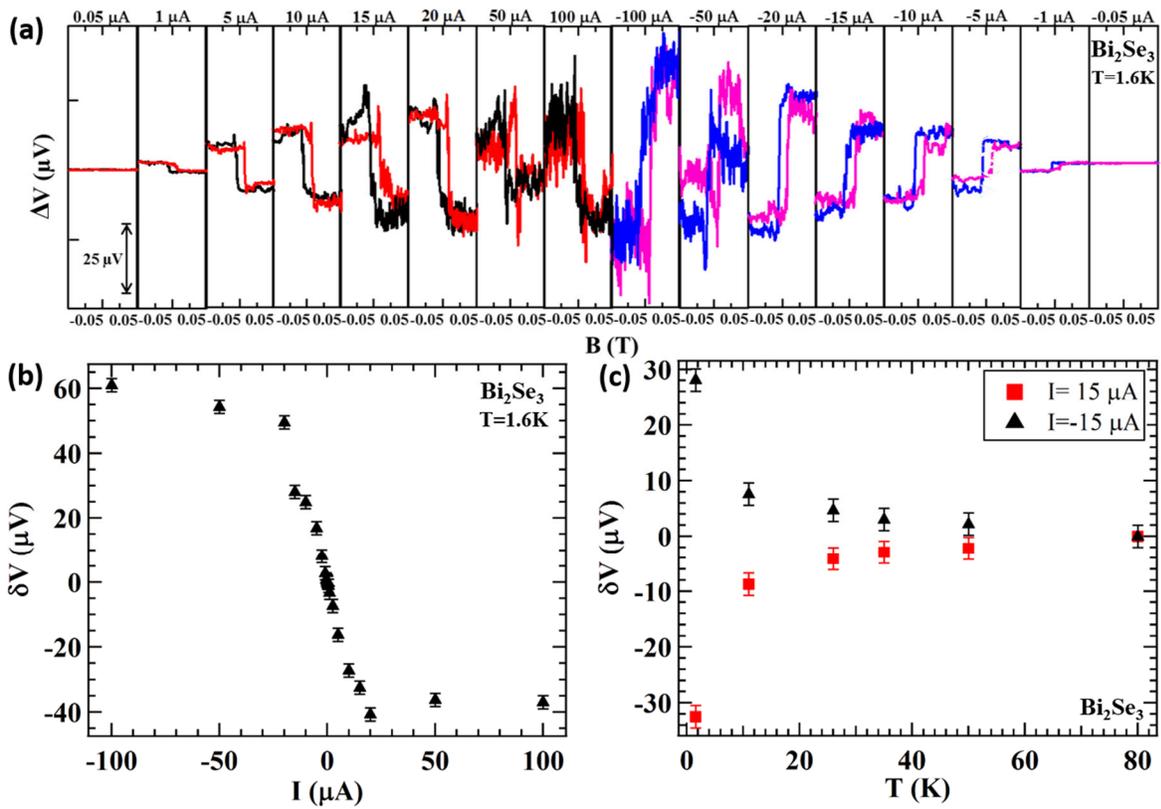

Figure 5 by Tian et al.

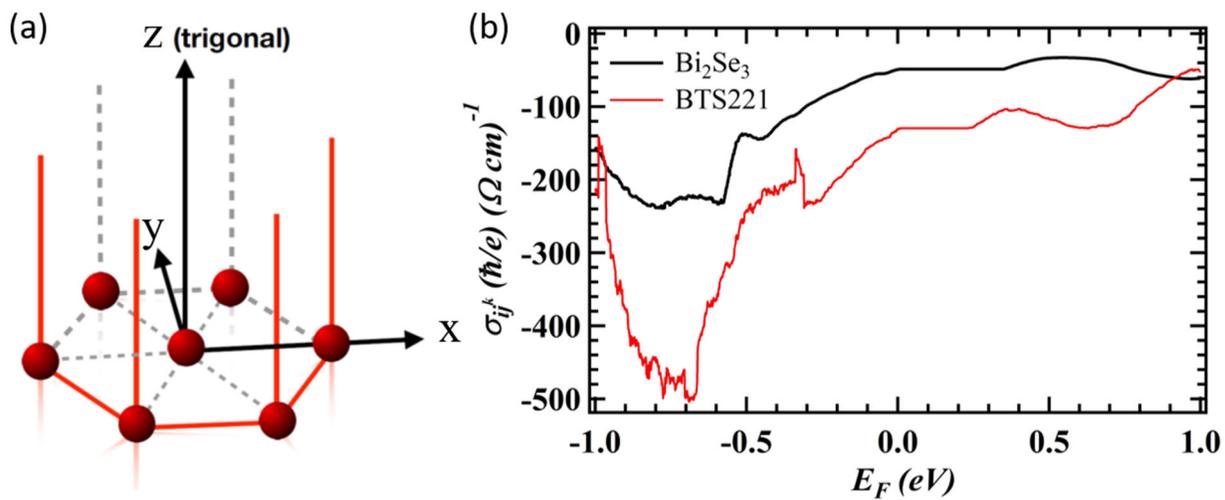

Figure 6 by Tian et al.